\documentclass[10pt,pra,aps,twocolumn,superscriptaddress]{revtex4-2}

\usepackage[utf8]{inputenc}
\usepackage{hyperref}
\hypersetup{colorlinks,allcolors=blue}
\usepackage{amsmath,revsymb,amssymb,mathtools}
\usepackage{graphicx}
\usepackage{bbold,comment}
\usepackage{braket,ulem}
\usepackage{booktabs}
\usepackage{float}
\usepackage{physics}
\usepackage{parskip}
\usepackage{multirow}
\usepackage{tabularx}
\usepackage{chemformula}
\usepackage{array}
\usepackage{pgf}
\usepackage{csquotes}

\newcolumntype{C}[1]{>{\centering\let\newline\\\arraybackslash\hspace{0pt}}m{#1}}
\newcolumntype{Y}{>{\centering\arraybackslash}X}

\newcommand{\til}{~}

\begin{document}
\title{Versatile Wavelength-Division Multiplexed Quantum Key Distribution Network Operating Simultaneously in the O and C Bands}
\author{Davide Scalcon}
\thanks{Affiliated to Toshiba Europe Ltd at the time of drafting.}
\affiliation{Dipartimento di Ingegneria dell'Informazione, Universit\`a degli Studi di Padova, via Gradenigo 6B, IT-35131 Padova, Italy}

\author{Matteo Padovan}
\affiliation{Dipartimento di Ingegneria dell'Informazione, Universit\`a degli Studi di Padova, via Gradenigo 6B, IT-35131 Padova, Italy}

\author{Paolo Villoresi}
\affiliation{Dipartimento di Ingegneria dell'Informazione, Universit\`a degli Studi di Padova, via Gradenigo 6B, IT-35131 Padova, Italy}
\affiliation{Padua Quantum Technologies Research Center, Universit\`a degli Studi di Padova, via Gradenigo 6B, IT-35131 Padova, Italy}

\author{Giuseppe Vallone}
\affiliation{Dipartimento di Ingegneria dell'Informazione, Universit\`a degli Studi di Padova, via Gradenigo 6B, IT-35131 Padova, Italy}
\affiliation{Padua Quantum Technologies Research Center, Universit\`a degli Studi di Padova, via Gradenigo 6B, IT-35131 Padova, Italy}

\author{Marco Avesani}
\email{marco.avesani@unipd.it}
\affiliation{Dipartimento di Ingegneria dell'Informazione, Universit\`a degli Studi di Padova, via Gradenigo 6B, IT-35131 Padova, Italy}

\begin{abstract}
    Ongoing technological progress is accelerating the commercial and global-scale deployment of Quantum Key Distribution (QKD). 
    Its ability to enable unconditionally secure communication is expected to be a key feature of future telecommunication networks, and practical demonstrations of QKD network implementations in real-world environments are crucial for ensuring reliable adoption. 
    In this work, we demonstrate a four-node photonic QKD network that employs versatile and cost-effective wavelength-division multiplexing across three transmitters in the O and C bands to simultaneously distribute quantum-secure keys among all nodes. 
    Specifically, the broadband central receiver node shares all optical and electronic decoding components, except for the single-photon detectors, across the three QKD links, significantly reducing system costs and enhancing compactness.
\end{abstract}

\maketitle
\section{Introduction}
Quantum Key Distribution (QKD) has emerged as a cornerstone of the second quantum revolution, driving the transition of quantum technologies from research to real-world applications. 
With the promise of information-theoretic security guaranteed by the laws of physics \cite{Bennett2014_BB84, Scarani2008, Pirandola2019rev, Sidhu2021}, QKD is at the center of numerous international initiatives focused on the development of a global quantum communication network \cite{Raymer_2019, Mehic2020, revQKDNets}. 
A key step toward this goal is to take advantage of the existing telecommunications infrastructure, which provides a natural pathway for deployment.

In recent years, research has been increasingly focused on improving the reliability, robustness, cost effectiveness, and scalability of photonic-based QKD systems \cite{Duligall_2006, Sibson:17, Xia:2019, Agnesi:20, Avesani:21, Avesani:22VSIX, scalcon2022cross, Picciariello2025}. 
Significant milestones have been achieved in both terrestrial and satellite-based QKD networks. 
Some examples include the Tokyo QKD network, which demonstrated real-world field tests of secure quantum communication \cite{Sasaki2011}, and Chinese effort on satellite QKD \cite{Liao2017, Chen2021, Li2025}.
In Europe, the SECOQC network in Vienna was one of the first large-scale QKD testbeds \cite{Peev2009}, while the Cambridge Quantum Network has contributed to integrating QKD into fiber-optic communication \cite{Dynes2019}.
At a broader level, the European Quantum Communication Infrastructure (EuroQCI) project aims to establish a pan-European quantum-secure network by interconnecting national QKD infrastructures \cite{EuroQCI,openqkd}.
For an exhaustive review of QKD networks, see \cite{revQKDNets}.
These advancements mark significant progress toward a future where quantum-secured communication becomes an integral part of our digital landscape.

To build scalable and cost-effective QKD networks, sharing resources via multiplexing is essential. 
While time-division multiplexing (TDM) can reduce detector counts using optical switches \cite{Picciariello2025}, it prevents simultaneous operation, thus reducing the overall key generation rate per user. 
Wavelength-division multiplexing (WDM) overcomes this limitation by allowing multiple QKD links to run in parallel. We highlight that this work focuses on WDM for multiple QKD links, a distinct challenge from co-propagating QKD with classical data channels at different wavelengths \cite{Bahrami2020,Peters_2009,Shao:25,Iqbal2025,Avesani:22VSIX, hajomer2025,Wang2015}.
Previous implementations of WDM-QKD networks have faced significant trade-offs. 
Many have relied on time-bin encoding \cite{Terhaar:23,Beutel:22,Yoshino:12}, but this approach's dependence on phase stability makes it inherently sensitive to wavelength. 
Consequently, the shared measurement hardware is only effective for channels operating in very close spectral proximity, typically within the same telecom band, which limits network versatility. 
An alternative approach used polarization-entangled photons \cite{Wengerowsky2018}; however, entanglement-based protocols typically yield lower secret key rates than prepare-and-measure schemes.
We overcome these limitations with an architecture based on polarization encoding, which allows for inherently broadband operation.
Specifically, we realized a four-node photonic quantum network, demonstrating a capability not previously shown in WDM-QKD systems with shared receivers. 
In our demonstration, three transmitters simultaneously exchange keys with a central trusted relay node using a single, shared measurement apparatus. 
Critically, our system concurrently serves channels operating in the widely separated 1310 nm (O-band) and 1550 nm (C-band) telecommunication windows. 
This wide cross-band versatility is achieved while maintaining a cost-effective design using standard, commercially available optics and requiring only one SPAD per channel. 
To the best of our knowledge, this work represents the first experimental demonstration of a prepare-and-measure WDM-QKD network capable of simultaneous, cross-band operation with a single shared receiver, thus presenting a practical and scalable path for integrating QKD into diverse fiber-optic infrastructures.

\begin{figure}
    \centering
    \includegraphics[width=0.95\linewidth]{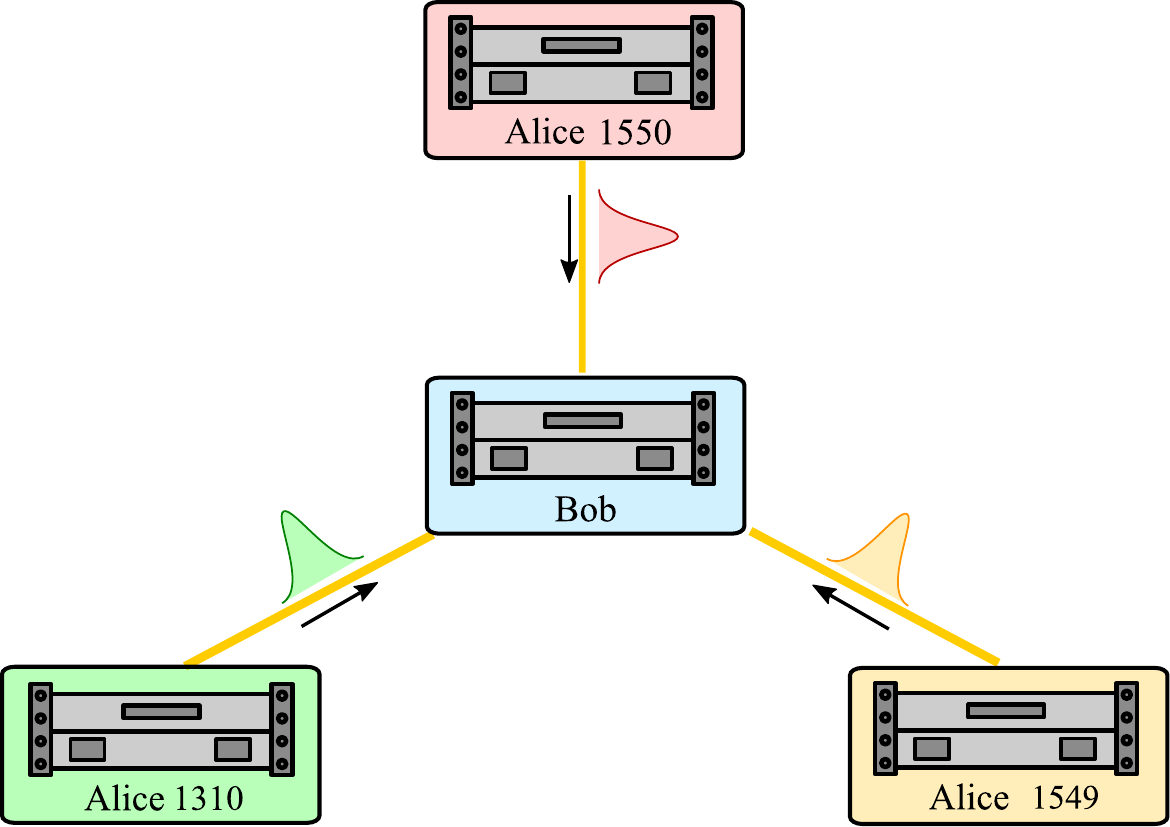}
    \caption{The star network implemented during the experiment.}
    \label{fig:star}
\end{figure}

\begin{figure*}
    \centering
    \includegraphics[width=.9\linewidth]{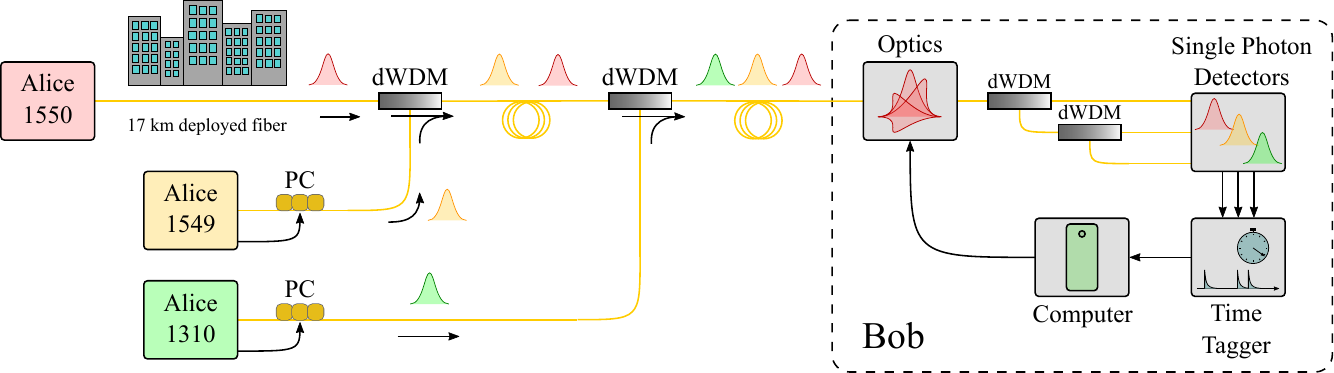}
    \caption{Scheme of the QKD network realized in the field trial.}
    \label{fig:infra}
\end{figure*}

\section{Methods}
\subsection{Infrastructure}

The QKD network we realized is shown in Figs.\ \ref{fig:star} and \ref{fig:infra}: it is a star configuration made up of three transmitters, called Alice$_{1550}$, Alice$_{1549}$, and Alice$_{1310}$, and one receiver, Bob.
Alice$_{1550}$ is located at the National Institute for Nuclear Physics (INFN) in Legnaro and is connected to Bob's location with a 17-km already deployed telecommunication optical fiber.

The three transmitters operate at three different wavelengths: 1550.12\til{nm} (CH34 of the DWDM ITU grid), 1549.32\til{nm} (CH35 of the DWDM ITU grid), and 1310\til{nm}, and their quantum channels are multiplexed into one single-mode optical fiber directed to Bob using both dense and coarse wavelength division multiplexers.
In particular, Alice$_{1549}$ and Alice$_{1310}$ are in the same laboratory as Bob and, together with Alice$_{1550}$, are connected to it through a spool of optical fiber with added extra losses.
The three quantum channels have an attenuation of approximately 14 dB, 12 dB, and 11 dB, for Alice$_{1550}$, Alice$_{1549}$ and Alice$_{1310}$, respectively.

The deployed systems implemented the efficient 3-state variant of the BB84 protocol \cite{Fung2006, Grunenfelder2018}. 
This variation of the original BB84 protocol enables some simplifications in the transmitter design and in the post-processing software.

\subsection{Transmitters}
\begin{figure}
    \centering
    \includegraphics[width=1\linewidth]{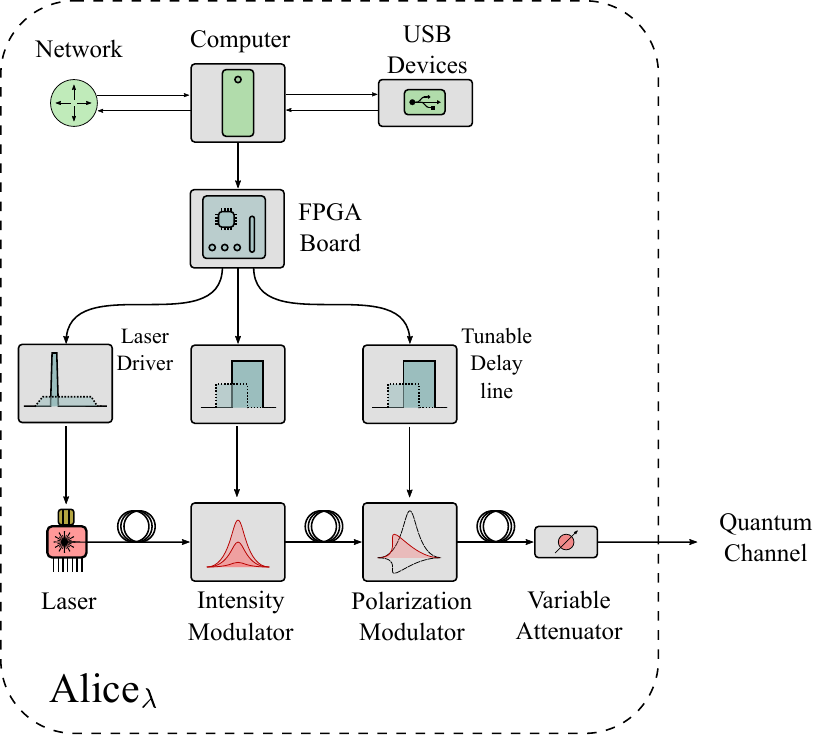}
    \caption{Scheme of the QKD transmitters used in the field trial.
    The same scheme is used in the three different
    Alice$_{1550}$, Alice$_{1549}$, and Alice$_{1310}$ transmitter}.
    \label{fig:alice_scheme}
\end{figure}

The three transmitters share a common architecture, depicted on the left in Fig. \ref{fig:alice_scheme}, consisting of a laser photon source, an intensity modulator, a polarization modulator, and control electronics and devices \cite{Picciariello2025,Lazzarin2024,Picciariello2023,Berra2023,Avesani:22VSIX,Avesani:21,Avesani2021b,Agnesi:20,Agnesi2019}.

The laser sources are either a \(1550\)~nm or a \(1310\)~nm gain-switched distributed feedback laser, generating pulses with a full-width-at-half-maximum (FWHM) of \(270\)~ps at a repetition rate of \(50\)~MHz. 
Specifically, one of the two \(1550\)~nm lasers is detuned to \(1549\)~nm by adjusting its operating temperature.

The intensity modulation is implemented using a fiber-optic Sagnac loop, with a 70:30 beam splitter (BS), a lithium niobate (\ch{LiNbO_3}) phase modulator, and a \(1\)m-long delay line\til\cite{Roberts2018}. 
This setup enables the decoy-state protocol by modulating the mean photon number of the transmitted pulse between two levels, \(\mu\) and \(\nu\), maintaining a ratio of \(\mu/\nu \approx 3.5\). 
Both the source and the intensity modulator consist entirely of polarization-maintaining (PM) fiber components.

The polarization modulation is performed using the iPOGNAC \cite{Avesani:21} for Alice\(_{1550}\) and Alice\(_{1310}\), while the POGNAC \cite{Agnesi2019} is used for Alice\(_{1549}\). 
Both configurations are based on a Sagnac interferometer that incorporates a lithium niobate (\ch{LiNbO_3}) phase modulator and a \(1\)m-long delay line. The incoming light is equally divided into clockwise and counterclockwise propagation modes within the loop. 
By leveraging the asymmetry of the interferometer and precisely controlling the modulator voltage and pulse timing, the scheme generates two orthogonal circular polarization states, \(\ket{R}\) and \(\ket{L}\). 
When no phase shift is applied, the resulting polarization state is diagonal, denoted as \(\ket{D}\).
We note that in the 
in the efficient 3-state variant of the BB84 protocol \cite{Fung2006, Grunenfelder2018}, it is not necessary to prepare the anti-diagonal state $\ket{A}$.
Finally, to achieve single-photon operation, the signal is attenuated using a variable optical attenuator (VOA) to \(\mu = 0.6\) and \(\nu = 0.17\) before transmission over the quantum channel.

In addition, Alice$_{1549}$ and Alice$_{1310}$ employ a polarization controller (PC) to compensate for the polarization transformation due to their optical fiber channel and to align their polarization reference system with the one of Alice$_{1550}$ and Bob. 
Before running the QKD protocol, an alignment phase is needed. 
First, Alice$_{1550}$ sends a sequence of known states, which the receiver uses to align the projection optics. 
This first phase ensures the receiver is measuring on two mutually unbiased basis.
Once the receiver is ready, the two other transmitters match their polarization reference frame using the polarization controller at their output.

The transmitter control system is a System-on-a-Chip (SoC) that integrates an FPGA and a CPU on a dedicated board (ZedBoard by Avnet). 
This SoC synchronizes the generation of signals that drive the laser and the two electro-optical phase modulators, following a real-time generated pseudo-random sequence.  
For a complete overview of the SoC architecture, refer to \cite{Stanco2021}. 
The post-processing, communication, and autonomous system control are managed by custom-designed software running on external PCs.

\subsection{Receiver}
\begin{figure*}
    \centering
    \includegraphics[width=.8\linewidth]{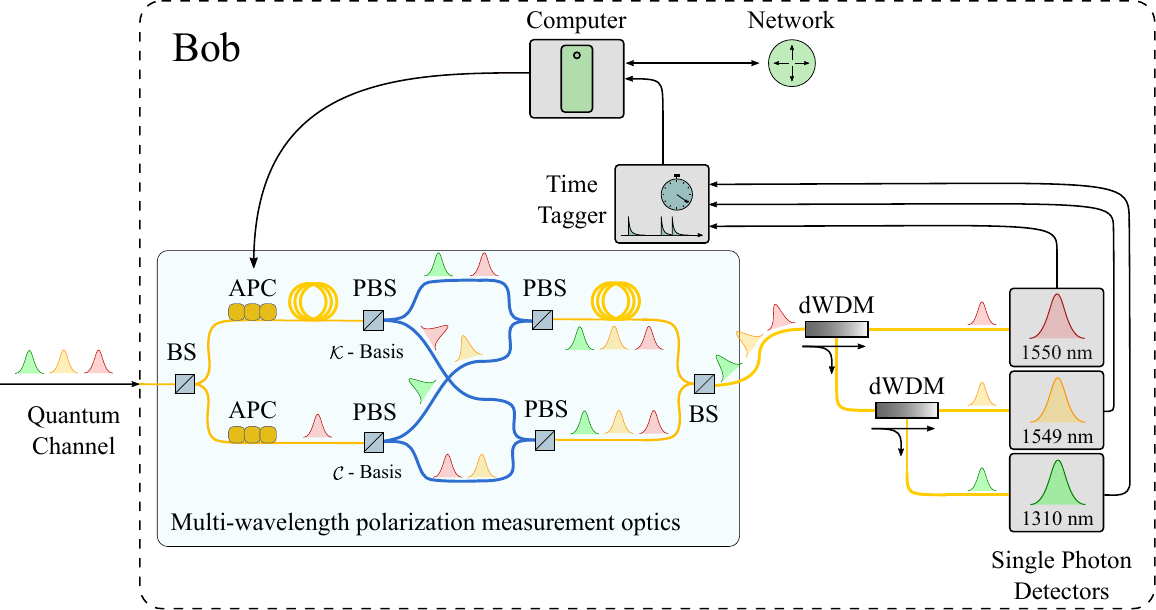}
    \caption{Details of the QKD receiver used in the field trial. Wavelength-division multiplexing was successfully implemented thanks to the measurement optics operating correctly at all three wavelengths: the depiction of more pulses inside the measurement apparatus than at the input is for illustrative purposes only and does not reflect the actual experimental conditions, where only one pulse is present per measurement round.}
    \label{fig:bob_scheme}
\end{figure*}

A key feature of the system is that on Bob's side, the measurement of the photon polarization is performed using a single broadband optical setup capable of operating at all three wavelengths. 
To further reduce the required hardware, this system, shown in Fig.\ \ref{fig:bob_scheme}, implements a polarization to time multiplexing by mapping the four polarizations $\ket L$, $\ket R$, $\ket D$ and $\ket A$ to four distinct time slots, allowing the use of only one detector per wavelength. 
Standard commercial fiber components such as PBS (polarizing beam splitter) and BS (beam splitter) are used to implement the polarization to time multiplexing mapping. 
Automatic Polarization Controllers (APCs) are used to determine the measurement basis, namely   $\mathcal K$-basis $\{\ket L,\ket R\}$ and   $\mathcal C$-basis $\{\ket D,\ket A\}$.
We note that the polarization measurement is inherently multi-wavelength since the used components (fibers, PBS, and BS) are broadband and they perform similarly with the three used wavelengths 1550nm, 1549nm and 1310m: as a result, a photon with given polarization $\ket L$, $\ket R$, $\ket D$ or $\ket A$ will be mapped to different time slots irrespectively of its wavelength.
This feature is the key novelty introduced in the present work:  three transmitters (the number can be extended to arbitrary number of dWDM channels)  Alice$_{1550}$, Alice$_{1549}$ and Alice$_{1310}$ can work
simultaneously with a single polarization measurement apparatus.

Although the polarization-time conversion can reduce the number of detectors, it is not essential, and the same scheme can be realized with 2 or 4 detectors, increasing the system performances. 
The photons are finally directed to the appropriate detectors using dedicated dense wavelength division demultiplexers, which introduces approximately \(1\) dB of additional losses per dWDM.
The detectors used are InGaAs/InP single-photon avalanche diodes (SPADs), specifically the PDM-IR model from Micro Photon Devices S.r.l., which provides a quantum efficiency of \(15\%\).
Photon arrival times are recorded using a time-to-digital converter (quTAU from qutools GmbH) and sent to Bob's PC for data processing. 
Time synchronization is achieved through the Qubit4Sync algorithm \cite{Calderaro2020}, which reconstructs the clock period based on detection events without requiring an external reference signal or additional fibers. 
This approach further minimizes the fiber's requirement to run the parallel QKD exchange.

\section{Results and Discussion}
\begin{figure}
    \centering
    \includegraphics[width=\linewidth]{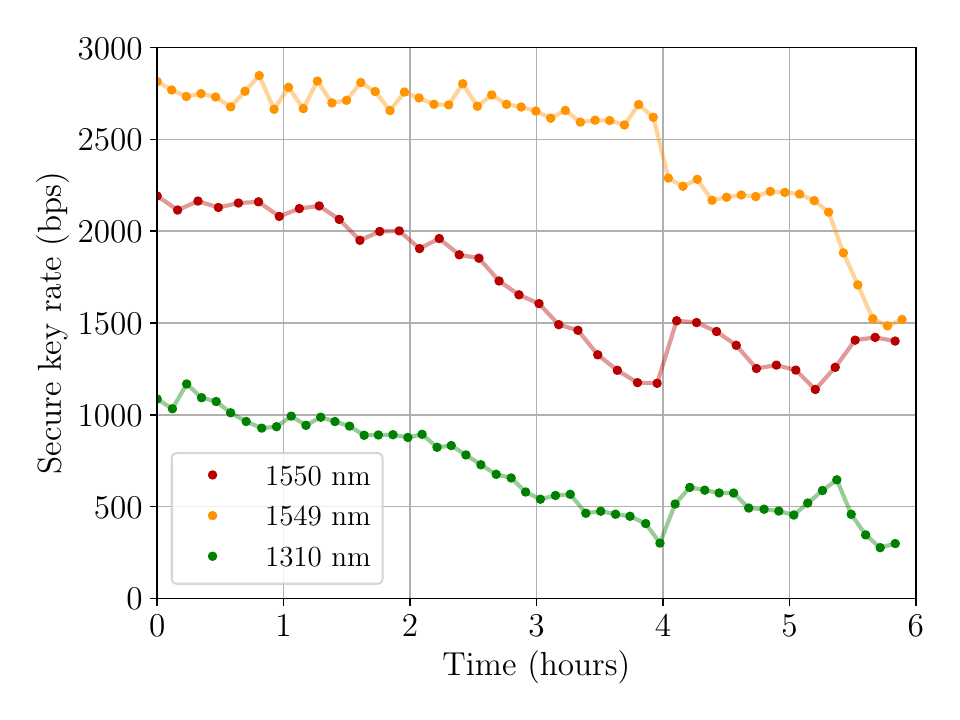}
    \caption{Secure key rate for the three parallel systems achieved during the experiments.}
    \label{fig:skr}
\end{figure}

\begin{figure}
    \centering
    \includegraphics[width=\linewidth]{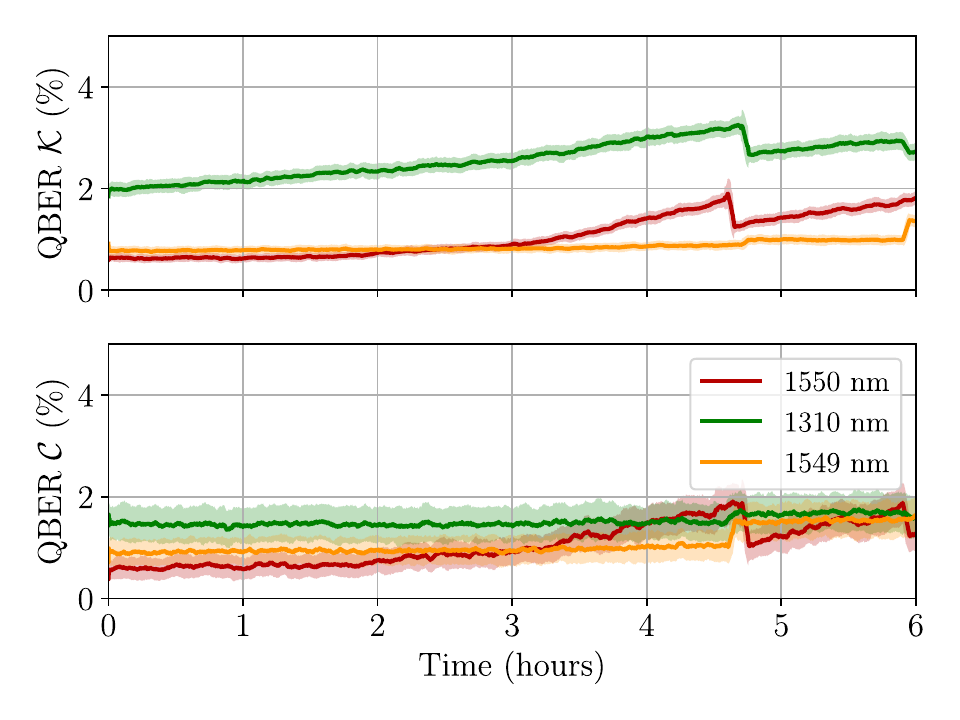}
    \caption{Quantum Bit Error Rate in the key and check basis for the three parallel systems measured by the post-processing during the experiments.}
    \label{fig:qbers}
\end{figure}

The three QKD links operated simultaneously for approximately 6 hours. 

The post-processing system operates in parallel with quantum transmission for each of the QKD nodes and delivers keys in real time.
The authentication of the classical channel is initially achieved with a pre-shared random sequence, which is later replaced by small portions of the QKD-generated keys as soon as they became available.
The error correction is implemented using a modified version of the Cascade algorithm \cite{Martinez-Mateo2015}, achieving a Shannon inefficiency of approximately \(1.05\).

The secret key generation followed the finite-size analysis presented in Ref. \cite{Rusca2018}, which is derived from Ref. \cite{Lim2014} and is robust against general attacks. 
This method is applied to sifted key blocks of \(5 \cdot 10^6\) bits in the key basis \(\mathcal{K}\), which are then shortened by the privacy amplification procedure to a final secret key length (SKL) given by:
\begin{equation}
 \mathrm{SKL} = s_{\mathcal{K},0} + s_{\mathcal{K},1}(1 - h(\phi_\mathcal{K})) - \lambda_{\rm EC} -\lambda_{\rm c} - \lambda_{\rm sec} \,.
\end{equation}
Here, \(s_{\mathcal{K},0}\) and \(s_{\mathcal{K},1}\) represent the lower bounds on the number of vacuum and single-photon detection events on the key basis, respectively. 
The term \(\phi_\mathcal{K}\) denotes the upper bound of the phase error rate for single-photon pulses, while \(h(\cdot)\) is the binary entropy function. 
The parameters \(\lambda_{\rm EC}\) and \(\lambda_{\rm c}\) correspond to the number of bits disclosed during error correction and correctness confirmation. 
Finally, \(\lambda_{\rm sec}\) accounts for the secrecy parameter and is defined as $\lambda_{\rm sec}=6 \log_2\left(\frac{19}{\epsilon_{\rm sec}}\right)$, where \(\epsilon_{\rm sec} = 10^{-10}\) represents the security parameter associated with the secrecy analysis.

With these post-processing parameters, we obtained the following results.  

The 1550 nm link measured an average QBER of 1.1\% on both bases and generated a total key of approximately 36.4 Mbits.  
The 1549 nm link measured an average QBER of 1\% in both bases, producing a total key of approximately 53.7 Mbits.  
The 1310 nm link measured an average QBER of 2.6\% in the $\mathcal{K}$-basis and 1.5\% in the $\mathcal{C}$-basis, resulting in a total key of approximately 15 Mbits.
This higher QBER, mainly due to the misalignment between Alice’s and Bob’s bases, explains the lower key rate compared to the other two links.
Figure~\ref{fig:skr} presents the secret key rate achieved during the experiments, in which the negative slope of the curves is due to the absence of active control to compensate for the polarization changes induced by the single mode fiber in the quantum channel for the 1549 nm and 1310 nm links.
The 1550 nm link implemented active control to maintain QBER values below 2\%.
The QBERs in the $\mathcal{K}$-basis and $\mathcal{C}$-basis as a function of time are shown in Figure~\ref{fig:qbers}.

\section{Conclusions}

In this work, we have demonstrated a highly versatile WDM-QKD network architecture that achieves simultaneous secure key distribution between multiple nodes on a single optical fiber. 
The principal innovation of our system is its ability to support concurrent operation of channels spanning the widely separated 1310 nm (O-band) and 1550 nm (C-band) telecommunication windows, all processed by a single, shared receiver.
The significance of this result lies in its practical implications for the deployment of quantum networks. 
By successfully multiplexing multiple QKD links onto a single fiber, our approach minimizes the need for dedicated fiber links per user, aligning QKD technology with the resource-efficient model of classical WDM networks. 
Crucially, the demonstrated broadband compatibility proves that a single network node can serve a heterogeneous mix of users and technologies without requiring separate, wavelength-specific hardware. This capability not only enhances the scalability of the network, but also significantly reduces the cost and complexity of the central receiving node, a key bottleneck in network expansion.
Looking ahead, our results pave the way for more sophisticated and integrated quantum communication systems. 
Immediate future work will focus on automating polarization alignment to ensure robust and long-term stability. Beyond this, the successful demonstration of multiband QKD on a single fiber opens a promising path toward full integration of parallel QKD links with classical data traffic (following the approach in \cite{Avesani:22VSIX}), allowing to exploit even further the existing telecommunication infrastructure.

\section*{Acknowledgments}
This work was supported by European Union's Horizon Europe research and innovation program under the project Quantum Secure Networks Partnership (QSNP), grant agreement No 101114043. 
Views and opinions expressed are however those of the authors only and do not necessarily reflect those of the European Union or European Commission-EU. 
Neither the European Union nor the granting authority can be held responsible for them.

\bibliography{biblio}

\end{document}